\documentclass[aip,
 cha,
 amsmath,
 amssymb,
 preprint,
 fleqn
]{revtex4-1}

\usepackage{bm}
\usepackage[dvipdfmx]{graphicx}
\usepackage[usenames]{color}
\usepackage[normalem]{ulem}

\graphicspath{{figs/}}
\begin{document}


\title{Phase-amplitude reduction of transient dynamics far from attractors for limit-cycling systems} 

\author{S. Shirasaka}
\email{Corresponding author: shirasaka.s.aa@m.titech.ac.jp}
\affiliation{Graduate School of Information Science and Engineering,
Tokyo Institute of Technology, O-okayama 2-12-1, Meguro, Tokyo 152-8552, Japan}

\author{W. Kurebayashi}
\affiliation{Faculty of Software and Information Technology, Aomori University, Kobata 2-3-1, Aomori, Aomori 030-0943, Japan}

\author{H. Nakao}
\affiliation{School of Engineering,
Tokyo Institute of Technology, O-okayama 2-12-1, Meguro, Tokyo 152-8552, Japan}

\date{\today}

\begin{abstract}
Phase reduction framework for limit-cycling systems based on isochrons
has been used as a powerful tool for analyzing rhythmic phenomena.
Recently, the notion of isostables, which complements the isochrons
by characterizing amplitudes of the system state, i.e., deviations from
the limit-cycle attractor, has been introduced to describe transient 
dynamics around the limit cycle
[Wilson and Moehlis, Phys. Rev. E {\bf 94}, 052213 (2016)].
In this study, we introduce a framework for a reduced phase-amplitude
description of transient dynamics of stable limit-cycling systems.
In contrast to the preceding study, the isostables are treated
in a fully consistent way with the Koopman operator analysis, which enables
us to avoid discontinuities of the isostables and to apply the 
framework to system states far from the limit cycle.
We also propose a new, convenient bi-orthogonalization method
to obtain the response functions of the amplitudes, which can be
interpreted as an extension of the adjoint covariant Lyapunov
vector to transient dynamics in limit-cycling systems. 
We illustrate the utility of the proposed reduction framework by
estimating optimal injection timing of external input
that efficiently suppresses deviations of the system state
from the limit cycle in a model of a biochemical oscillator.
\end{abstract}

\pacs{
05.45.-a 
}

\maketitle 


\begin{quotation}
The phase reduction theory provides a general framework to simplify a complex,
multi-dimensional limit-cycling system describing a stable rhythmic activity to
a one-dimensional phase equation evolving on a circle~\cite{winfree2001geometry,kuramoto2012chemical,hoppensteadt2012weakly,ermentrout2010mathematical,nakao2016phase,ashwin2016mathematical}. 
It has been successfully used to understand synchronization phenomena of weakly interacting rhythmic elements in physical, chemical, biological and engineered systems~\cite{winfree2001geometry,kuramoto2012chemical,hoppensteadt2012weakly,ermentrout2010mathematical,nakao2016phase,ashwin2016mathematical,pikovsky2003synchronization,strogatz2005theoretical,tinsley2012chimera,tass2007phase,schultheiss2011phase,dorfler2013synchronization}.  
Methods to optimize and control synchronization of rhythmic elements have also
been developed by using the phase reduction framework~\cite{moehlis2006optimal,kiss2007engineering,harada2010optimal,zlotnik2013optimal,zlotnik2016phase}. 
However, to describe the system dynamics far from the limit cycle,
amplitude degrees of freedom should be taken into account.
In this study, by extending preceding studies, we propose a phase-amplitude reduction
framework that is applicable to transient dynamics far from the limit cycle.
\end{quotation}

\section{Introduction} \label{sec. intro}
The roles of amplitude degrees of freedom in limit-cycling systems,
which represent deviations of the system states from the limit-cycle attractor
and are eliminated in the phase-reduction framework, have been extensively
studied because they are rich sources of intriguing oscillator dynamics at individual~\cite{pikovsky2003synchronization,yoshimura2008phase,goldobin2010dynamics,wedgwood2013phase,hitczenko2013poincare,mauroy2014global,ashwin2016mathematical} and ensemble~\cite{kuramoto2012chemical,matthews1991dynamics,cross1993pattern,aranson2002world,pikovsky2003synchronization,nakao2009diffusion,koseska2013oscillation,ashwin2016mathematical} levels.
In most studies, however, the analysis is restricted to the vicinity of a supercritical Hopf bifurcation, where a simple normal form 
(Stuart-Landau equation) of the oscillator dynamics is available~\cite{guckenheimer2013nonlinear,strogatz2014nonlinear}. 
Some other studies use moving orthonormal frames along the limit cycle
to define the amplitudes of the oscillator~\cite{wedgwood2013phase,hitczenko2013poincare,ashwin2016mathematical},
which allow the quantitative study of the amplitude dynamics
of oscillators far from bifurcation points. However, in general,
those amplitude variables interact nonlinearly with each other,
which hinders simplification of the system description. 
Thus, it is highly desirable to establish a framework for
a quantitative reduced description of limit-cycling systems
applicable to transient dynamics far from the limit cycle.
Such a framework would facilitate in-depth studies of the roles of amplitude
degrees of freedom of limit-cycling systems in realistic settings. 

The key idea in the phase reduction is assigning the same phase value
to the set of initial conditions that share the same asymptotic behavior. 
These sets of identical phase values are called {\it isochrons}~\cite{winfree2001geometry,kuramoto2012chemical,hoppensteadt2012weakly,ermentrout2010mathematical,nakao2016phase,pikovsky2003synchronization}. 
Analogously, in a recent work~\cite{mauroy2013isostables}, the notion of {\it isostables}
is introduced by identifying the initial conditions that share the same relaxation property, i.e., the same decay rate toward the attractor. 
It has also been shown~\cite{mauroy2013isostables} that the isochrons and isostables can be
understood from a unified point of view of the spectral properties of the
{\it Koopman} ({\it composition}) {\it operator}~\cite{budivsic2012applied}. 
For each characteristic decay rate of the system state toward the attractor,
a set of isostables representing an amplitude degree of freedom can be introduced,
which is independent from the phase and the other amplitude degrees of freedom. 
By retaining a small number of amplitude variables representing
dominant (slowly-decaying) part of the transient dynamics, reduced description
of the system dynamics can be derived. 
The Koopman operator has attracted broad interest recently, because it is closely related to a rapidly developing 
data-driven approach to complex nonlinear systems, called the dynamic mode decomposition~\cite{budivsic2012applied,rowley2009spectral,schmid2010dynamic,williams2015data,erichson2016compressed,doi:10.1080/14697688.2016.1170194,proctor2016dynamic}. 

Amplitude reduction frameworks for a system near a stable equilibrium
based on isostables have been established for multi-dimensional~\cite{mauroy2013isostables,mauroy2014converging,wilson2015extending} and infinite-dimensional systems~\cite{wilson2016isostable} and have been used to formulate optimal control problems of 
moving the system state toward the equilibrium~\cite{mauroy2013isostables,wilson2015extending,wilson2016isostable}. 
Recently, Wilson and Moehlis~\cite{PhysRevE.94.052213} have extended
the isostable reduction framework to limit-cycling systems.
However, the isostables introduced in their work have discontinuities on one leaf of the isochrons. 
To avoid this problem, it is assumed in Ref.~\cite{PhysRevE.94.052213} that the system evolves
in a close-enough neighborhood of the limit cycle so that the discontinuities are negligible,
and the amplitude response to perturbation in their reduced system involves the first
order response evaluated only on the limit cycle. 
Therefore, their analysis is essentially equivalent to deriving a decoupled linear system 
preserving spectral properties of the original system in a vicinity of the limit-cycle attractor 
(called {\it kinematically similar} system in terms of {\it Lyapunov transformations}~\cite{lyapunov1992general,adrianova1995introduction,colonius2014dynamical}) 
by making use of covariant properties of {\it adjoint covariant Lyapunov vectors}~\cite{kuptsov2012theory} 
(also called adjoint Floquet vectors~\cite{Traversa2015} or dual Lyapunov vectors~\cite{pikovsky2015}). 
A method to analyze response functions of decoupled phase and amplitude variables in limit-cycling systems, 
which is based on the {\it Lie symmetries} formalism and is valid far from the attractors, has also been proposed~\cite{guillamon2009computational,castejon2013phase}. 
However, the latter analysis is limited to two-dimensional dynamical systems and naive application of the method proposed in Ref.~\cite{castejon2013phase}, that is, solving {\it adjoint equations} to calculate the response functions, can yield flawed results numerically, as we discuss in this paper. 

In this study, we introduce a phase-amplitude reduction framework to describe transient dynamics of stable limit-cycle oscillators, which is applicable to high-dimensional dynamics far from the limit-cycle attractor.
We propose a systematic bi-orthogonalization method to numerically estimate the
fundamental quantities for the reduction accurately, i.e., the first order response functions
of the phase and amplitudes to perturbations along a given trajectory,
which is not necessarily the limit cycle itself.
These response functions can be interpreted as an extension of the adjoint covariant Lyapunov vectors to transient dynamics. 
We illustrate the utility of the proposed framework by estimating optimal injection timing of external input that realizes maximal suppression of the most persistent (least decaying) amplitude degree of freedom.

This paper is organized as follows: in Sec.~\ref{sec. theory}, phase and amplitudes in limit-cycling systems are introduced using the Koopman operator theory. 
In Sec.~\ref{sec. mtd}, the phase-amplitude reduction framework for limit-cycling systems is introduced and the bi-orthogonalization method to obtain their response properties is developed. 
In Sec.~\ref{sec. example}, the theory is illustrated by analyzing the phase-amplitude response properties of a minimal chemical kinetic model of an oscillatory genetic circuit. 
Also, the optimal injection timing problem is introduced and analyzed. 
Section~\ref{sec. concl} summarizes the results. 
\section{Phase, amplitudes and the Koopman operator} \label{sec. theory}
We consider a $N$-dimensional autonomous dynamical system 
\begin{equation}
	\dot{\bm X} = {\bm F}({\bm X}), \hspace{1cm} {\bm X} \in \mathbb{R}^N, \label{eq. system}
\end{equation}
where ${\bm X}(t)$ is a system state and ${\bm F}({\bm X})$ is a vector field. 
Suppose the system (\ref{eq. system}) has a periodic orbit $\chi: {\bm X}_0(t)$ with period $T$. 
Let $\phi : \mathbb{R}\times \mathbb{R}^N \to \mathbb{R}^N$ denote the flow induced by Eq.~(\ref{eq. system}), i.e., $\phi(t, {\bm X})$ is the solution of Eq.~(\ref{eq. system}) at the time $t$ with the initial condition ${\bm X}$ at $t = 0$. 

The stability of the periodic orbit $\chi$ is characterized by the {\it characteristic multipliers}~\cite{guckenheimer2013nonlinear} $\Lambda_i \ (i=1,\cdots,N)$, which are the eigenvalues of the time-$T$ flow linearized around a point ${\bm X}_0(t_*)$ on the orbit $\chi$ (also called the {\it monodromy matrix}): ${\bf M}({\bm X}_0(t_*)) = \partial \phi(T, {\bm X})/\partial {\bm X}|_{{\bm X}={\bm X}_0(t_*)}$. 
When the relation $1 = \Lambda_1 > |\Lambda_2| \ge \cdots \ge |\Lambda_N|$ holds, the periodic orbit $\chi$ is a stable limit cycle. 
For simplicity, we hereafter assume that the Floquet multipliers $\Lambda_i$ are positive, real, and simple.
Extension to the case with complex conjugate multipliers can be performed in a parallel way to the analysis of stable equilibria~\cite{mauroy2013isostables,wilson2016isostable}. 
We consider dynamics of the system in the basin of attraction ${\mathcal B} \subset \mathbb{R}^N$ of the stable limit cycle $\chi$. 

The Koopman operator $U^t$ is a linear operator that describes the evolution of a function defined on the phase space, called an observable $f: \mathbb{R}^N \to \mathbb{C}$. It is defined as $U^t f({\bm X}) = f \circ \phi(t,{\bm X})$, where $\circ$ represents composition of functions. 
The operator $U^t$ has eigenfunctions~\cite{mezic2013analysis,lan2013linearization} $s_i({\bm X}) \ (i=1,\cdots, N)$ associated with eigenvalues $\lambda_i \ (i=1,\cdots, N)$, that is, 
\begin{equation}
	U^ts_i({\bm X}) = e^{\lambda_i t}s_i({\bm X}), 
\end{equation}
where $\lambda_1 = \sqrt{-1} \omega$, $\omega \equiv 2\pi/T$, and $\lambda_i = \mathrm{log}(\Lambda_i)/T \ (i=2,\cdots,N)$. 
The eigenvalues correspond to the {\it characteristic exponents} of the limit cycle $\chi$~\cite{guckenheimer2013nonlinear}, hence they reflect the spectral property of the limit-cycling system. 

We hereafter assume that the vector field ${\bm F}$ is twice continuously differentiable so that the continuously differentiable eigenfunctions $s_i$ exist on the whole basin of attraction~\cite{lan2013linearization}, and we further assume the gradients of $s_i$ are {\it Lipschitz continuous} on $\mathcal{B}$, which is required for the perturbative analysis. 
Note that a {\it non-resonant} analyticity of ${\bm F}$, which holds generically in practical situations, is sufficient for the Lipschitz continuity, because this assures that $s_i$ is analytic. 

Let us introduce amplitudes of the system state ${\bm X}$ by $r_i({\bm X}) \equiv \mathrm{Re}(s_i({\bm X})) \ (i=2,\cdots,N)$, where $\mathrm{Re}(z)$ is the real part of a complex number $z$. Because
\begin{equation}
	U^{\Delta t}r_i({\bm X}) = \mathrm{Re}(s_i(\phi(\Delta t, {\bm X}))) = e^{\lambda_i \Delta t}r_i({\bm X}), 
\end{equation}
each $r_i$ obeys
\begin{equation}
	\dot{r_i}({\bm X}) = \lim_{\Delta t \to 0}{\dfrac{U^{\Delta t}r_i({\bm X})-r_i({\bm X})}{\Delta t}} = \lambda_i r_i. \label{eq. amp}
\end{equation}
We can also introduce a phase of ${\bm X}$ by $\theta({\bm X}) \equiv \mathrm{arg}(s_1({\bm X}))$, where $\mathrm{arg}(z)$ is the argument of $z$, whose range is defined as the interval $[0,2\pi)$. Because $\lambda_1 = \sqrt{-1} \omega$, $\theta$ obeys
\begin{equation}
	\dot{\theta}({\bm X}) = \omega. \label{eq. phase}
\end{equation}
This definition of the phase coincides with that of the {\em asymptotic phase} used in the conventional phase reduction theory~\cite{winfree2001geometry,kuramoto2012chemical,hoppensteadt2012weakly,ermentrout2010mathematical,nakao2016phase,ashwin2016mathematical}. 
Therefore, level sets of $\theta$ provide isochrons.  
Analogously, isostables are defined as level sets of $|r_i|$. 
Note that the linear form (\ref{eq. amp},\ref{eq. phase}), which is valid in the entire basin of attraction~\cite{lan2013linearization}, 
is not necessarily derived by the perturbative power-series approach based on the {\it Poincar{\'e}-Dulac normal form theory} and its extensions~\cite{guckenheimer2013nonlinear,shilnikov1998methods,gaeta2002poincare,wiggins2003introduction,sanders2007averaging}. 
Hence we do not assume the non-resonance condition usually required for a complete linearization in the Poincar{\'e}-Dulac type scheme. 
See Sec.~3.2 of Lan and Mezi{\'c}'s work~\cite{lan2013linearization} for an example with resonance that can be linearized by using Koopman eigenfunctions including non-analytic (trans)monomials. 

Because the sign of $r_i$ is neglected, each isostable is composed of two connected components corresponding to $+r_i$ and $-r_i$. 
These connected components of isostables, associated with one of the exponents $\lambda_i$, foliate the basin of attraction of the limit cycle, and each leaf of this foliation provides a level set of the amplitude associated with the exponent.  
From Eq.~(\ref{eq. amp}), we can see that initial conditions on the same isostable share the same decay rate toward the limit cycle. 
These phase and amplitudes defined above evolve independently under linear time invariant dynamics and thus provide simple
description of the dynamics around the limit cycle.

Here, we note that the amplitudes can also be defined as $\tilde{r}_i({\bm X}) \equiv |s_i({\bm X})|$, as in the preceding study~\cite{mauroy2013isostables}. 
However, this definition makes a coordinate transformation ${\bm X} \mapsto (\theta,\tilde{r}_2,\cdots,\tilde{r}_N)^{\dag}$ ($\dag$ denotes transpose) non-invertible, 
i.e., its inversion can be multi-valued in some region. 
The phase-amplitude expression may suffer from this ambiguity, particularly when we apply perturbations to the system. 
Therefore, we adopt the definition $r_i({\bm X}) \equiv \mathrm{Re}(s_i({\bm X}))$ in this study. 

\section{Reduction framework and a method to calculate the response functions of the phase and amplitudes} \label{sec. mtd}

Suppose that perturbation $\epsilon {\bm p}(t)$, where $\epsilon > 0$ characterizes its magnitude, is introduced to the oscillator (\ref{eq. system}) as
\begin{equation}
	\dot{\bm X} = {\bm F}({\bm X}) + \epsilon {\bm p}(t). \label{eq. ptsyst}
\end{equation}
We denote a coordinate transformation ${\bm X} \mapsto \Theta$ by ${\bm X} = {\bm h}(\Theta)$, where $\Theta = (\theta,r_2,\cdots,r_N)^
\dag$. 
In this phase-amplitudes coordinate, the perturbed system (\ref{eq. ptsyst}) takes the following form: 
\begin{align}
	\dot{\theta} &= \omega + \epsilon \nabla \theta({\bm h}(\Theta)) \cdot {\bm p}(t),  \label{eq. ptphase} \\
	\dot{r_i} &= \lambda_i r_i + \epsilon \nabla r_i({\bm h}(\Theta)) \cdot {\bm p}(t), \quad (i=2,\cdots, N),
	\label{eq. ptamp} 
\end{align}
where $\nabla$ represents gradient and $\cdot$ is a dot product. 
Consider a solution $\chi^*:{\bm X}^*(t)$ of the unperturbed system (\ref{eq. system}) with an initial condition ${\bm X}^*(0)$ taken arbitrarily in the basin of attraction $\mathcal{B}$, 
and let $\chi_p^*: {\bm X}_p^*(t)$ be a solution of the perturbed system (\ref{eq. ptsyst}) with the same initial condition ${\bm X}_p^*(0) = {\bm X}^*(0)$ as the unperturbed system.    
As is known in a regular perturbation theory~\cite{sanders2007averaging,coddington1955theory,hoppensteadt2000analysis,chicone2006ordinary,teschl2012ordinary}, we can show by the {\it Gr{\"o}nwall-Bellman inequality} that the magnitude of the error $||{\bm X}_p^*(t) - {\bm X}^*(t)||$, where $||\cdot ||$ denotes the Euclidean norm, is bounded by $b\epsilon(e^{at}-1)/a$, where $a$ and $b$ are positive constants. 
This means that ${\bm X}_p^*(t)$ is in a neighborhood of radius $\epsilon$ of ${\bm X}^*(t)$ within a finite time interval of length $O(1)$. 
We here emphasize that this does not imply the breakdown of the continuous dependence of the solutions on $\epsilon$ within a specific, fixed finite time interval 
(as long as the unperturbed solution exists on an entire half line, which is the case here). 
In fact, once we fix an arbitrary large finite length interval $[0,T_{\rm f}]$, we can consider ${\bm X}_p^*(t)$ is in a neighborhood of 
radius $\epsilon$ of ${\bm X}^*(t)$ on this interval by taking appropriately small $\epsilon$, because $T_{\rm f}$ is independent of $\epsilon$, and this is sufficient for our argument. 
The fact that the length of this interval is $O(1)$ means that the convergence of ${\bm X}_p^*(t)$ to ${\bm X}^*(t)$ is {\it non-uniform} on an $\epsilon$-dependent 
interval $[0,\epsilon^\beta)$ for any $\beta < 0$, i.e., the limiting passages $t \to \epsilon^\beta$ and $\epsilon \to +0$ cannot be interchanged. 
This does not affect our analysis in this study, because no asymptotic properties of the perturbed dynamics are discussed. 
In this interval, we can expand the gradients using the Lipschitz continuity as $\nabla \theta({\bm h}(\Theta)) = \nabla \theta({\bm X}^*(t)) + O(\epsilon)$ and $\nabla r_i({\bm h}(\Theta)) = \nabla r_i({\bm X}^*(t)) + O(\epsilon)$ in Eqs.~(\ref{eq. ptphase},\ref{eq. ptamp}). 
Thus, we can approximate Eqs.~(\ref{eq. ptphase},\ref{eq. ptamp}) as 
\begin{align}
	\dot{\theta} &= \omega + \epsilon \nabla \theta({\bm X}^*(t)) \cdot {\bm p}(t),  \label{eq. trphase} \\
	\dot{r_i} &= \lambda_i r_i + \epsilon \nabla r_i({\bm X}^*(t)) \cdot {\bm p}(t), \quad (i=2,\cdots, N), \label{eq. tramp} 
\end{align}
by neglecting the terms of order $\epsilon^2$. 

These equations are completely decoupled from each other and we can adopt combinations of these $N$ equations (\ref{eq. trphase},\ref{eq. tramp}) as a reduced form of the system dynamics in the close-enough neighborhood of the transient trajectory $\chi_*$.
In most cases, the first $K$ equations of (\ref{eq. trphase},\ref{eq. tramp}) for some $K (\ll N)$ are of interest, because they describe relatively persistent, slowly decaying modes. Hereafter, we discuss a method to obtain the reduced $K$ equations. 
The phase and amplitude response functions to perturbation, $\nabla \theta({\bm X}^*(t))$ and $\nabla r_i({\bm X}^*(t))$, are the fundamental quantities for the proposed reduction framework. 

First, we evaluate the gradients on the periodic orbit $\chi$.
Consider an initial condition slightly deviated from the periodic orbit, ${\bm h}_p \equiv {\bm h}(\Theta_1) + \delta{\bm x}$, where we defined $\Theta_1 = (\theta, 0, \cdots, 0)^{\dag}$. 
Then
\begin{equation}
	U^Tr_i({\bm h}_p) = e^{\lambda_i T}r_i({\bm h}(\Theta_1)+\delta{\bm x}). \label{eq. Tmap1}
\end{equation}
Using the time-$T$ flow, we can also express $U^Tr_i({\bm h}_p) $ as
\begin{align}
	U^Tr_i({\bm h}_p) = r_i({\bm h}(\Theta_1)+{\bf M}({\bm h}(\Theta_1))\delta{\bm x} + O(||\delta{\bm x}||^2)). 
	\label{eq. Tmap2}
\end{align}
Equating the RHSs of Eqs.~(\ref{eq. Tmap1},\ref{eq. Tmap2}), Taylor expanding $r_i$ around ${\bm h}(\Theta_1)$, considering that $r_i({\bm h}(\Theta_1))=0$ and that the direction of $\delta {\bm x}$ is arbitrary and taking the limit $||\delta {\bm x}|| \to 0$, we can show that
\begin{equation}
	\nabla r_i^{\dag}({\bm h}(\Theta_1)){\bf M}({\bm h}(\Theta_1)) = e^{\lambda_i T}\nabla r_i^{\dag}({\bm h}(\Theta_1)). \label{eq. monoadjr} 
\end{equation}
Similarly, we obtain 
\begin{equation}
\nabla \theta^{\dag}({\bm h}(\Theta_1)){\bf M}({\bm h}(\Theta_1)) = \nabla \theta^{\dag}({\bm h}(\Theta_1)). \label{eq. monoadjt}
\end{equation}
Thus, the gradient vectors of the phase and amplitudes evaluated on $\chi$ are left eigenvectors of the monodromy matrix, which are called the adjoint covariant Lyapunov vectors~\cite{kuptsov2012theory,Traversa2015,pikovsky2015}.
These vectors can be numerically obtained by the {\it QR-decomposition} based methods~\cite{kuptsov2012theory,pikovsky2015} or by the {\it spectral dichotomy} approaches~\cite{froyland2013computing,Thorsten2016computing}. 

Next, we seek the equations for the gradients of the phase and amplitudes on the transient trajectory $\chi^*: {\bm X}^*(t)$. 
Here, we introduce logarithmic amplitudes $\psi_i({\bm X}) \equiv \mathrm{log}(|r_i({\bm X})|) \ (i=2,\cdots,N)$ 
in order to make the following treatment of the gradients of the amplitudes simple and parallel with the standard arguments in the conventional phase reduction theory. 
For convenience of notation, let $\psi_1({\bm X}) = \theta({\bm X}).$  
In the following, we evaluate the gradient vectors of $\psi_i$, whose directions coincide with those of $\theta$ and $r_i$. 
The gradients $\nabla \theta$ and $\nabla r_i$ can be calculated from $\nabla \psi_i$ by rescaling, where the following normalization conditions should be satisfied:
\begin{align}
	\nabla r_i({\bm X}^*(t))\cdot {\bm F}({\bm X}^*(t)) = \lambda_i r_i, \label{eq. normr} \\
	\nabla \theta({\bm X}^*(t))\cdot {\bm F}({\bm X}^*(t)) = \omega. \label{eq. normt}
\end{align}
These normalization conditions are equivalent to Eqs.~(\ref{eq. amp},\ref{eq. phase}).

We can derive adjoint equations for the gradients by using the same argument as the conventional derivation of the adjoint equation for the phase response curves, given by Brown et al.~\cite{brown2004phase}.
	It is well known that an infinitesimal error $\delta {\bm x}(0)$ introduced at $t=0$ between two unperturbed solutions ${\bm X}^*(t)+\delta {\bm x}(t)$ and ${\bm X}^*(t)$ satisfies the {\it variational equation}~\cite{guckenheimer2013nonlinear,shilnikov1998methods,sanders2007averaging,coddington1955theory,chicone2006ordinary,teschl2012ordinary} ${\rm d}(\delta{\bm x}(t))/{\rm d}t = {\rm D}{\bm F}({\bm X}^*(t))\delta{\bm x}(t)$. 
	Because each logarithmic amplitude $\psi_i$ increases constantly as $\dot{\psi_i}({\bm X}(t)) = \nabla \psi_i({\bm X}(t)) \cdot \dot{\bm X}(t) = \lambda_i$ in the absence of perturbation, the error in the logarithmic amplitude coordinate $\psi_i({\bm X}^*(t)+\delta {\bm x}(t)) - \psi_i({\bm X}^*(t)) = \nabla \psi_i({\bm X}^*(t)) \cdot \delta {\bm x}(t)$ should be independent of time, i.e., ${\rm d}(\nabla \psi_i({\bm X}^*(t)) \cdot \delta {\bm x}(t))/{\rm d}t = 0$. This yields 
\begin{align}
	\frac{{\rm d} \nabla \psi_i({\bm X}^*(t))}{{\rm d}t}\cdot \delta {\bm x}(t) &= - \nabla \psi_i ({\bm X}^*(t))\cdot \frac{{\rm d}(\delta {\bm x}(t))}{{\rm d}t}  \nonumber \\
											   &= - \nabla \psi_i ({\bm X}^*(t))\cdot {\rm D}{\bm F}({\bm X}^*(t))\delta{\bm x}(t)  \nonumber \\
		     &= -{\rm D} {\bm F}^{\dag}({\bm X}^* (t))\nabla \psi_i ({\bm X}^*(t)) \cdot \delta{\bm x}(t).  \label{eq. deradj}
\end{align}
Here we used the variational equation and the definition of the adjoint matrix. 
We can take $N$ linearly independent initial errors $\delta {\bm x}_i(0) = \epsilon'{\bm e}_i$, where $0<\epsilon' \ll 1$ and ${\bm e}_i$ is the $i$th unit vector and define the {\it fundamental solution matrix} ${\bf L}(t)$ of the variational equation as ${\bf L}(t) = (\delta {\bm x}_1(t),\delta {\bm x}_2(t),\cdots,\delta {\bm x}_N(t))$. 
The sign of the determinant of the fundamental solution matrix, called the {\it Wronskian}, is time-invariant due to {\it Liouville's trace formula}~\cite{adrianova1995introduction,colonius2014dynamical,wiggins2003introduction,coddington1955theory,chicone2006ordinary,teschl2012ordinary}. 
Because ${\rm det}({\bf L}(0)) = (\epsilon')^N>0$, we obtain ${\rm det}({\bf L}(t))>0$ for all $t$, and thus the fundamental solution matrix is always invertible. 
Consider a matrix form of the Eq.~({\ref{eq. deradj}}), $({\rm d}(\nabla \psi_i({\bm X}^*(t)) / {\rm d}t) {\bf L}(t) = -{\rm D} {\bm F}^{\dag}({\bm X}^* (t))\nabla \psi_i ({\bm X}^*(t)) {\bf L}(t)$. 
We can eliminate ${\bf L}(t)$ by multiplying its inverse from the right side on both sides of this equation. 
Therefore, 
\begin{equation}
	\frac{{\rm d} \nabla \psi_i({\bm X}^*(t))}{{\rm d}t} = -{\rm D} {\bm F}^{\dag}({\bm X}^* (t))\nabla \psi_i ({\bm X}^*(t))  \label{eq. adjeq}
\end{equation}
should hold. Note that this equation should be solved with an appropriate end condition. 
Here, we can approximately take the end condition of Eq.~(\ref{eq. adjeq}) as $\nabla \psi_i({\bm X}^*(\tau)) \parallel \nabla r_i ({\bm h}(\Theta_1))|_{\theta = \theta_*}$ for some $t=\tau$ and $\theta=\theta_*$, because the gradient field $\nabla r_i({\bm X})$ is continuous and the transient trajectory eventually converges to the limit cycle. 
The adjoint tangent {\it propagator} $\mathcal{G}(t_1,t_2) \equiv {\bf N}(t_2){\bf N}^{-1}(t_1)$, where ${\bf N}(t)$ is a fundamental solution matrix of the linear system given by Eq.~(\ref{eq. adjeq}), maps $\nabla \psi_i ({\bm X}^*(t_1))$ to $\nabla \psi_i ({\bm X}^*(t_2))$. 
Thus, $\nabla \theta ({\bm X}^*(t_2)) \parallel \mathcal{G}(t_1,t_2)\nabla \theta ({\bm X}^*(t_1))$ and $\nabla r_i ({\bm X}^*(t_2)) \parallel \mathcal{G}(t_1,t_2)\nabla r_i ({\bm X}^*(t_1))$ hold.
Therefore, the gradient vectors of the phase and amplitudes are covariant with respect to the action of the propagator $\mathcal{G}$ and they can be interpreted as an extension of the adjoint covariant Lyapunov vectors to transient regimes (note that the adjoint covariant Lyapunov vectors evaluated on the limit cycle, given by Eqs.~(\ref{eq. monoadjr},\ref{eq. monoadjt}), are covariant w.r.t. the action of the adjoint of the monodromy matrix, which is the one period (time-$T$) propagator). 

In the numerical estimation of $\nabla \theta$ (or $\nabla \psi_1$), a standard method is to integrate the adjoint equation backward in time, while renormalizing $\nabla \theta$ occasionally so that the normalization condition (\ref{eq. normt}) is satisfied~\cite{ermentrout2010mathematical}.
	This is because $\nabla \theta$ corresponds to the neutrally stable component ($\mbox{Re} (\lambda_1) = 0$) while other components have negative growth rates  ($\lambda_{2, ..., N} < 0$).
However, in the present case, naive backward integration does not provide correct results for the amplitudes, $\psi_{2, ..., N}$, because vector components caused by numerical errors in the relatively (backward-in-time) unstable covariant subspaces accumulate.
Therefore, we have to develop a method to subtract them off. 
Note that the standard QR-decomposition based methods~\cite{kuptsov2012theory,pikovsky2015} to obtain the covariant subspace require the ergodicity of the underlying dynamical process, hence they cannot be directly applied to the process far from attractors, and that the spectral dichotomy techniques~\cite{froyland2013computing,Thorsten2016computing} to evaluate them may not work well near the left boundary of the time evolution (See Secs.~2.6 and Sec.~2.7 of H\"{u}ls's work~\cite{Thorsten2016computing}).

To develop a numerical method, we introduce dual vectors ${\bm \gamma}_i$ of $\nabla \psi_i$ that are bi-orthogonal to $\nabla \psi_j$ as
\begin{equation}
	{\bm \gamma}_i({\bm X}^*(t))\cdot \nabla \psi_j({\bm X}^*(t)) = \delta_{ij}, \label{eq. dual}
\end{equation}
where $\delta_{ij}$ is the Kronecker delta. 
By using ${\bm \gamma}_i({\bm X}^*(t))$, we can subtract the vector component in the covariant subspace $\nabla \psi_i({\bm X}^*(t))$ from the solution ${\bm z}(t)$ of Eq.~(\ref{eq. adjeq}), which is given by projecting ${\bm z}(t)$ onto this subspace as
\begin{equation}
	({\bm \gamma}_i({\bm X}^*(t))\cdot {\bm z}(t))\nabla \psi_i({\bm X}^*(t)). \label{eq. proj}
\end{equation}
Differentiating Eq.~(\ref{eq. dual}) by $t$, we obtain $(\dot{{\bm \gamma}}_i({\bm X}^*(t)) - D{\bm F}({\bm X}^*(t)){\bm \gamma}_i({\bm X}^*(t)))\cdot \nabla \psi_j ({\bm X}^*(t)) = 0$. 
The sign of the Wronskian of Eq.~(\ref{eq. adjeq}) is time-invariant due to Liouville's trace formula. 
By using this fact and linear independence of the left eigenvectors of the monodromy matrix, we can show linear independence of $\{\nabla \psi_i ({\bm X)} \}_{i=1}^N$ for every point ${\bm X}$ in the whole basin of attraction $\mathcal{B}$. 
Thus, we obtain
\begin{equation}
	\dot{{\bm \gamma}}_i({\bm X}^*(t)) = D{\bm F}({\bm X}^*(t)){\bm \gamma}_i({\bm X}^*(t)). \label{eq. coveq}
\end{equation}
The vectors ${\bm \gamma}_i$ are covariant w.r.t. the action of the propagator ${\mathcal F}(= ({\mathcal G}^{\dag})^{-1})$ of the linear system (\ref{eq. coveq}), hence they can be seen as covariant Lyapunov vectors extended to transient regimes. 
The relative stability relation of covariant subspace of Eq.~(\ref{eq. coveq}) forward-in-time coincides with that of Eq.~(\ref{eq. adjeq}) backward-in-time.  
In order to subtract unstable components using the projection (\ref{eq. proj}), the system (\ref{eq. coveq}) should be solved forward-in-time with an approximate initial condition ${\bm \gamma}_i({\bm X}^*(0))$. 
The vectors $\{\nabla \psi_i ({\bm X^*(0))} \}_{i=1}^N$ can be approximated by direct numerical simulation of the dynamics, using the {\it Fourier averages} and the {\it generalized Laplace averages}~\cite{mauroy2012use,mezic2013analysis} (See Appendix~\ref{sec. gla} for details). 
Then, ${\bm \gamma}_i({\bm X}^*(0))$ can be obtained by using the bi-orthogonality relation (\ref{eq. dual}). 

Now, we introduce a bi-orthogonalization method to obtain the response functions of the phase and amplitudes up to the $K$th unstable mode. 
The procedure is as follows: 
(a) evaluate the adjoint Lyapunov vectors on the limit cycle $\chi$ and the characteristic exponents, 
(b) calculate $\{{\bm \gamma}_i({\bm X}^*(0))\}_{i=1}^K$ from $\{\nabla \psi_i ({\bm X^*(0))} \}_{i=1}^N$ obtained by direct numerical simulation using the bi-orthogonality relation (\ref{eq. dual}), 
(c) obtain $\nabla \psi_1 ({\bm X}^*(t))$ by backward integration of Eq.~(\ref{eq. adjeq}), 
(d) obtain ${\bm \gamma}_1({\bm X}^*(t))$ by forward integration of Eq.~(\ref{eq. coveq}), 
(e) obtain $\nabla \psi_2 ({\bm X}^*(t))$ by backward integration of Eq.~(\ref{eq. adjeq})
while subtracting relatively unstable mode $\nabla \psi_1 ({\bm X}^*(t))$ by the projection (\ref{eq. proj}), 
(f) obtain ${\bm \gamma}_2 ({\bm X}^*(t))$ by the forward integration of Eq.~(\ref{eq. coveq}) while
subtracting relatively unstable mode ${\bm \gamma}_1({\bm X}^*(t))$ by the projection 
\begin{equation}
	(\nabla \psi_i({\bm X}^*(t))\cdot {\bm y}(t)){\bm \gamma}_i({\bm X}^*(t)), \label{eq. covproj}
\end{equation}
where ${\bm y}(t)$ is a solution of Eq.~(\ref{eq. coveq}), 
(g) perform (e) and (f) consecutively to obtain $\{\nabla \psi_i ({\bm X^*(t))} \}_{i=3}^K$ and $\{{\bm \gamma}_i({\bm X}^*(t))\}_{i=3}^K$ (note that all relatively unstable modes should be subtracted during integration),   
(h) obtain $\nabla \theta$ and $\nabla r_i \ (i=2,\cdots,K)$ using the normalization conditions (\ref{eq. normr},\ref{eq. normt}), where $r_i({\bm X}^*(t))$ on the transient orbit $\chi^*$ is evaluated using Eq.~(\ref{eq. amp}) with the initial condition $r_i({\bm X}^*(0))$, which is calculated in (b) by the direct numerical simulation. 

This method has a significant computational advantages in evaluating the response functions.
To calculate response functions $\{\nabla \psi_i \}_{i=1}^K$ at $m$ points on the transient orbit $\chi^*$, it is necessary to repeat long-time evolution $mK(N+1)$ times if we evaluate them directly by the direct numerical simulation. 
In contrast, we need only $K(N+1)+2K$ times long-time evolution in the proposed bi-orthogonalization method. 

\section{Examples} \label{sec. example}
\begin{figure*}
	\includegraphics[width=\textwidth]{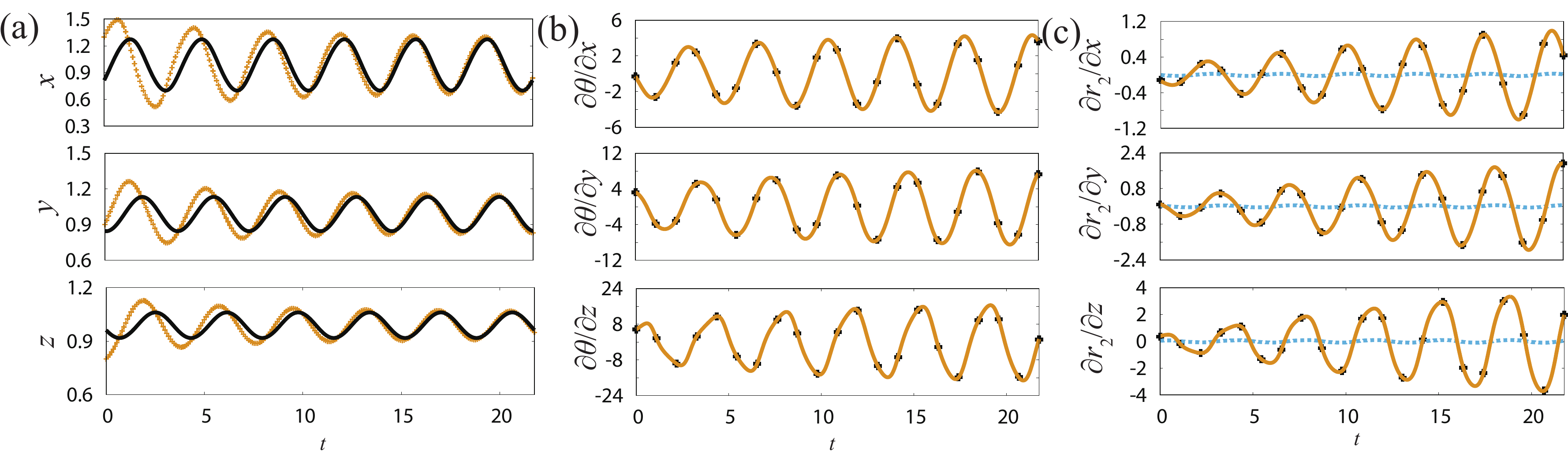}
	\caption{The Goodwin model. 
		(a) The stable periodic solution of the model (lines) and the transient solution ${\bm X}^*(t)$ (plus signs). 
		(b) Three components of the phase response function $\nabla \theta({\bm X}^*(t))$ obtained by the direct numerical simulation (plus signs) and by the backward integration of the adjoint equation (lines).
		(c) Three components of the second amplitude response function $\nabla r_2({\bm X}^*(t))$ obtained by the direct numerical simulation (plus signs), the naive backward integration method (blue dashed lines) and by the proposed bi-orthogonalization method (yellow lines). 
		They are all normalized using the condition (\ref{eq. normr}), and the results obtained by the direct numerical simulation are appropriately bi-orthogonalized to satisfy the duality relation (\ref{eq. dual}). 
}
	\label{fig. goodwin}
\end{figure*}

As an example, we analyze the Goodwin model, a minimal chemical kinetic model of an oscillatory genetic circuit~\cite{goodwin1965oscillatory,gonze2013goodwin}. 
The Goodwin model has a three-dimensional state ${\bm X} = (x, y, z)^\dag \in \mathbb{R}^3$. 
The state variables $x,y$, and $z$ can be interpreted as concentrations of a given clock mRNA, the corresponding protein, and a transcriptional inhibitor, respectively. 
We use a simple dimensionless form of the Goodwin model~\cite{woller2014goodwin}, 
\begin{align*}
	&\dot{x} = \frac{\alpha}{1+z^n} - x, \nonumber \\
	&\dot{y} = x - y, \nonumber \\
	&\dot{z} = y-z. \nonumber
\end{align*}
The parameters are set as $\alpha = 1.8$ and $n = 20$.
Figure~\ref{fig. goodwin}(a) shows the stable periodic solution of the model. 
The period and Lyapunov exponents are estimated as $T = 3.63$, $\lambda_2 = -0.0766$, and $\lambda_3 = -2.92$. 
We consider a transient solution ${\bm X}^*(t)$ with an initial condition ${\bm X}^*(0) = (1.30,0.900,0.800)^{\dag}$. 
Figure~\ref{fig. goodwin}(a) shows the transient solution.  
We set the end time $\tau = 63.0$ for the backward integration in the following calculation. 

In Fig.~\ref{fig. goodwin}(b), the phase response function $\nabla \theta({\bm X}^*(t))$ obtained by the backward integration of the adjoint equation (\ref{eq. adjeq}) is compared with the result of the direct numerical simulations.  
The results agree well, hence, along this transient solution ${\bm X}^*(t)$, $\nabla \theta({\bm X}^*(t))$ can always be considered as the most unstable covariant subspace. 

Figure~\ref{fig. goodwin}(c) shows the amplitude response functions $\nabla r_2({\bm X}^*(t))$, which is obtained by the proposed bi-orthogonalization method, by naive backward integration method, and by direct numerical simulations. 
All results are normalized using the condition (\ref{eq. normr}). 
Note here that, in the close-enough neighborhood of the limit cycle orbit $\chi$, the vectors $\nabla r_2({\bm X}^*(t))$ and ${\bm F}({\bm X}^*(t))$ are nearly normal. 
Hence, the normalization procedure using (\ref{eq. normr}) is very sensitive to tiny change in their directions. 
Therefore, not only the normalization condition (\ref{eq. normr}) but the duality relation (\ref{eq. dual}) must be carefully imposed on the results of the direct numerical simulation in order to make a reasonable comparison with those of the other methods. 
The results obtained by the naive backward integration considerably deviates from those obtained by direct numerical simulations, while those obtained by the proposed bi-orthogonalization method are in good agreement. 

Next, we illustrate the utility of the reduced amplitude equation (\ref{eq. tramp}) by estimating the optimal injection timing of weak external input to suppress the most persistent component $r_2$ of the amplitudes. 
We apply a transient control input $\epsilon{\bm p}(t)$ of a fixed waveform ${\bm w}$ and a fixed duration $\tau_*$, i.e., ${\bm p}(t)= {\bm w}(t-s)$ where ${\bm w}(\cdot)$ is nonzero only on $[0,\tau_*]$ and the time $s$ determines the injection timing of the input. 
In the spirit of Mauroy's preceding study~\cite{mauroy2014converging}, we introduce a finite-horizon optimal control problem of minimizing the amplitude $|r_2|$ at a given time $T_e$. 
This control problem can be formulated as follows: find the injection timing $s_*$ such that
\begin{equation}
	s_* = \mathrm{argmin}_{s\in \mathcal{I}_{\sigma}}|r_2({\bm X}^*_p(T_e))|, \label{eq. opt}
\end{equation}
where $\mathcal{I}_{\sigma} \equiv [0,T_e-\tau_*]$ and ${\bm X}^*_p(t)$ is the solution of Eq.~(\ref{eq. ptsyst}). 
When the magnitude of the input $\epsilon$ is sufficiently small, the evolution of the amplitude $r_2$ is approximated by the reduced equation (\ref{eq. tramp}). 
Then, using an analytical solution of the linear one-dimensional non-homogeneous differential equation (\ref{eq. tramp}) of $r_2$, the optimal control problem (\ref{eq. opt}) can be approximated to the problem of finding $s_*$ such that 
\begin{equation}
	\mathrm{sgn}(r_2({\bm X}^*(0)))\int_0^{T_e}{{\bm p}(t)\cdot \nabla r_2({\bm X}^*(t))e^{\lambda_2(T_e-t)}\mathrm{d}t} \label{eq. apx_opt}
\end{equation}
is minimized. 

\begin{figure}
	\includegraphics[width=0.5\textwidth]{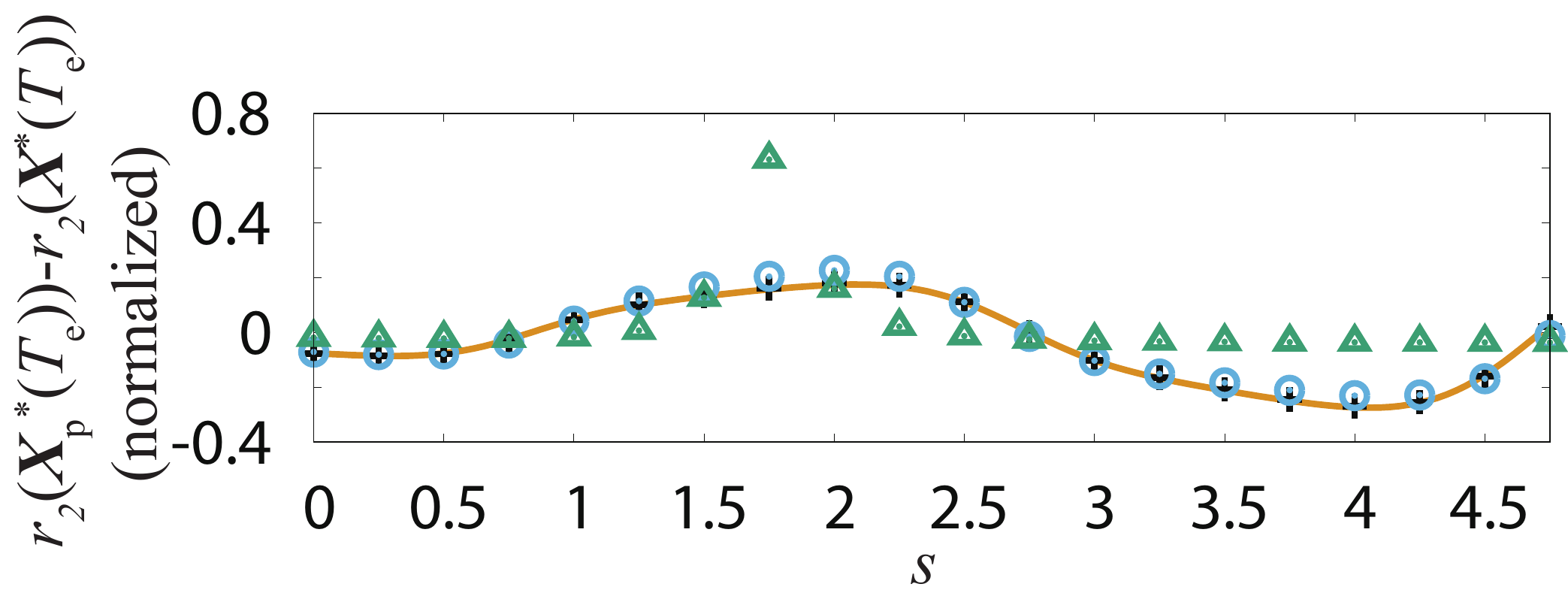}
	\caption{Optimal control problem for the Goodwin model. 
		Effect of the control input on the amplitude at a given time $r_2({\bm X}_p^*(T_e))$, obtained by an analytical solution of the reduced amplitude equation (line) and by the direct numerical simulations for 20 different injection timings for three different magnitudes of the input: $\epsilon = 0.01$ (black plus signs), $\epsilon = 0.1$ (blue circles) and $\epsilon = 1.0$ (green triangles). 
The results are normalized so that the $l_2$ norms of the waveforms evaluated using the 20 discrete time points are the same. 
}
	\label{fig. opt_contr}
\end{figure}

Figure~\ref{fig. opt_contr} shows the effect of the control input on the amplitude $r_2({\bm X}_p^*(T_e))$ at time $T_e = 5$. 
The control input is assumed as ${\bm w}(t) = (0,0,-1)^{\dag}$ and $\tau_* = 0.25$. 
The results obtained by the analytical solution of the reduced amplitude equation (\ref{eq. tramp}) is compared with the result of direct numerical simulations, showing good agreement for sufficiently weak input ($\epsilon = 0.01,0.1$). 
This verifies the validity of the approximate amplitude equation in the present situation. 
Thus, the optimal injection timing of sufficiently weak input can be theoretically predicted using the formula (\ref{eq. apx_opt}), because it is essentially equivalent to solving the approximate amplitude equation (\ref{eq. tramp}) directly. 
In this case, the initial value of the amplitude is negative, i.e., $r_2({\bm X}^*(0)) < 0$. 
Hence, the optimal injection timing $s_*$ of the sufficiently weak input can be estimated by finding the maximum of the waveform in Fig.~\ref{fig. opt_contr}, which gives $s_* = 2.08$ in this particular case. 
Finally, we note that when the magnitude becomes large $(\epsilon = 1.0)$, the approximation (\ref{eq. tramp}) fails and then the results considerably deviate from each other. 

\section{Conclusion} \label{sec. concl}
We formulated a phase-amplitude reduction framework for stable limit-cycling systems, which can be applied to 
transient dynamical regimes far from attractors in high-dimensional systems.
We also developed a bi-orthogonalization method for numerical estimation of the response function of 
the phase and amplitudes, which provides accurate phase-amplitude response functions.
As an application, we illustrated that the response functions accurately predicts the optimal injection timing of external input which efficiently suppress deviations from attractors. 
The proposed theory would be useful in analyzing and controlling response properties of high-dimensional rhythmic systems. 

\begin{acknowledgments}
We thank Yoshiyuki Yamaguchi for useful comments on this work. We are also grateful to anonymous reviewers for helpful suggestions and comments. 
S. S. acknowledges financial support from Japan Society for the Promotion of Science (JSPS) KAKENHI Grant No.~15J12045. W. K. acknowledges financial support from JSPS KAKENHI Grant No.~16K16125. H. N. acknowledges financial support from JSPS KAKENHI Grants No.~16H01538 and No.~16K13847. 
\end{acknowledgments}

\appendix

\section{The Fourier averages and the generalized Laplace averages}\label{sec. gla}
In this section, we introduce methods to obtain the phase and amplitudes by direct numerical simulation of the dynamics. 

The phase variable $\theta({\bm X})$ is evaluated as $\theta({\bm X}) = \mathrm{arg}(f_{\lambda_1}^*({\bm X}))$, where the Fourier average~\cite{mauroy2012use} $f_{\lambda_1}^*({\bm X})$ of an observable $f$ is given by
\begin{equation}
	f_{\lambda_1}^*({\bm X}) = \lim_{s \to \infty}{\dfrac{1}{s}\int_0^s{f\circ\phi(t,{\bm X})e^{-\lambda_1 t}\mathrm{d}t}}. 
\end{equation}
The amplitude variable $r_i({\bm X})$ is obtained by $r_i({\bm X}) = \mathrm{Re}(f_{\lambda_i}^*({\bm X}))$, where the generalized Laplace average~\cite{mezic2013analysis} $f_{\lambda_i}^*({\bm X})$ of $f$ is given by 
\begin{equation}
	f_{\lambda_i}^*({\bm X}) = \lim_{s \to \infty}{\dfrac{1}{s}\int_0^s{\left[f\circ\phi(t,{\bm X}) - \bar{f} - \sum_{k=1}^{i-1}{f_{\lambda_k}^*({\bm X})e^{\lambda_k t}}\right] e^{-\lambda_i t}}\mathrm{d}t }, 
\end{equation}
where $\bar{f}$ is an averaged observable along the periodic orbit $\chi$: $\bar{f} = (1/T)\int_0^T{f\circ\phi(t,{\bm X}_0(t_*))\mathrm{d}t}$. 

We can simplify the generalized Laplace averages using convenient observables $g_i \ (i=2, \cdots, N)$ defined as 
\begin{equation}
	g_i({\bm X}) = \nabla r_i({\bm X}_0(\theta_*)) \cdot ({\bm X} - {\bm X}_0(\theta_*)),
\end{equation}
where $\theta_* = \theta({\bm X})$. 
Here, the adjoint covariant Lyapunov vectors $\nabla r_i({\bm X}_0(\theta_*))$ are normalized so that they are dual to the unitized covariant Lyapunov vectors ${\bm \gamma}_i({\bm X}_0(\theta_*))$. 
Each of these observables evolves with its corresponding characteristic exponent asymptotically, because, in the close-enough neighborhood of the periodic orbit $\chi$, $g_i$ coincides with the $i$th amplitude variable $r_i$. 
Hence, we can show that $\bar{g}_i = 0$ and $(g_i)_{\lambda_k}^*({\bm X}) = 0 \ (k=1,\cdots,i-1)$ for any ${\bm X}$ in the basin of attraction $\mathcal{B}$. 
Thus, we can replace the generalized Laplace average with the Laplace average: 
\begin{equation}
	(g_i)_{\lambda_i}^*({\bm X}) = \lim_{s \to \infty}{\dfrac{1}{s}\int_0^s{g_i\circ\phi(t,{\bm X})e^{-\lambda_i t}}\mathrm{d}t }. 
\end{equation}
\providecommand{\noopsort}[1]{}\providecommand{\singleletter}[1]{#1}%

\end{document}